\documentstyle[preprint,aps]{revtex}
\begin{document}
\draft
\title{Two Model-Independent Results for the Momentum Dependence of
$\rho$-$\omega$ Mixing\\
\begin{flushright}
ADP-95-35/T189 \\
nucl-th/9506024
\end{flushright}}
\author{Kim Maltman\cite{byline}}
\address{Department of Mathematics and Statistics, York University,
4700 Keele St., \\ North York, Ontario, Canada M3J 1P3}
\date{\today}
\maketitle
\begin{abstract}
Two model-independent results on the momentum-dependence of $\rho$-$\omega$
mixing are described.  First, an explicit choice of interpolating
fields for the vector mesons is displayed for which both the
mixing in the propagator and the isospin-breaking at the nucleon-vector
meson vertices (and hence also the one-vector-meson-exchange
contribution to NN charge symmetry breaking) vanish identically
at $q^2=0$.  Second, it is shown,
using the constraints of unitarity and analyticity on the spectral
function of the vector meson propagator, that there is no
possible choice of interpolating fields for the $\rho^0$, $\omega^0$
mesons such that, with the $\rho\omega$ element of the
propagator defined by $\Delta^{\rho\omega}_{\mu\nu}(q^2)=
(g_{\mu\nu}-q_\mu q_\nu /q^2)$$\theta (q^2)/(q^2-m^2_\rho )(q^2-
m^2_\omega )$, $\theta (q^2)$ is independent of momentum.  It
follows that the standard treatment of charge symmetry breaking
in few-body systems cannot be interpreted as arising from any
realizable effective meson-baryon Lagrangian and must, therefore,
be considered purely phenomenological in content.
\end{abstract}

In standard meson-exchange models of few-body systems, isospin-breaking
meson-meson mixing plays an important role in generating contributions to
few-body charge-symmetry-violating (CSV)
observables.  Among these contributions, those associated with
$\rho$-$\omega$ mixing have, traditionally, been thought to be
rather well-determined, the mixing matrix element being taken
(under the somewhat questionable assumption of the absence of direct
$\omega^0\rightarrow\pi\pi$ contributions) to be
directly measured in the region of the $\rho$-$\omega$ interference
shoulder in $e^+e^-\rightarrow\pi^+\pi^-$~.  Using
this ``extracted'' value (i.e. assuming no ``direct''
$\omega^0\rightarrow\pi\pi$ coupling, where $\omega^0$
is the pure isospin zero $\omega$ state) one obtains significant
contributions to a number of observables, in particular,
the bulk of the non-Coulombic A=3 binding energy difference,
significant contributions to the $np$ asymmetry at $183$ MeV and
non-negligible contributions to the difference of $nn$ and $pp$
scattering lengths and the $np$ asymmetry at $477$ MeV
\onlinecite{ref1,ref2,ref3,ref4,ref5,ref6,ref7,ref8,ref9}.
This phenomenological success has, however, been recently called
into question by the suggestion that such mixing matrix elements
must, in general, be expected to be rather momentum-dependent
\onlinecite{ref10,ref11,ref12,ref13,ref14,ref15,ref16,ref17,ref18,ref19,ref20,ref21,ref22}.
If this were, indeed, the case then, even
assuming the interference in $e^+e^-\rightarrow\pi^+\pi^-$
in the vicinity of $q^2\sim m_\omega^2$ were to be dominated by
the $\rho$-$\omega$ mixing contribution,
the experimental input would not determine
the value of the mixing for $q^2<0$, where it is needed
in few-body CSV calculations.

A significant problem, which has considerably complicated
the discussion of this issue in the literature, is the dependence of
off-shell Green functions (such as the off-shell propagator)
on the choice of interpolating fields.  As is well-known, there
is no unique choice of fields to represent, eg., the
$\rho$, $\omega$ mesons:  given a particular choice $\{\rho ,\omega\}$
and a corresponding effective Lagrangian,
$L_{\rm eff}[\rho ,\omega ,\cdots ]$ (where $\cdots$ represents
all other fields), one may define
$\rho =\rho^\prime F(\rho^\prime )$ and $\omega =\omega^\prime
G(\omega^\prime )$, with $F(0)=G(0)=1$ and
$L^\prime _{\rm eff}[\rho^\prime ,\omega^\prime ,\cdots ]\equiv$
$L_{\rm eff}[\rho^\prime F(\rho^\prime ), \omega^\prime G(\omega^\prime ),
\cdots ]$.  For any such field redefinition, the
$\{\rho ,\omega ,L_{\rm eff}[\rho ,\omega ,\cdots ]\}$ and
$\{\rho^\prime ,\omega^\prime ,L_{\rm eff}[\rho^\prime ,
\omega^\prime ,\cdots ]\}$ theories have exactly
the same $S$-matrix elements\onlinecite{ref23,ref24} and
hence are physically equivalent.
The Green functions of the two theories, however, are not, in general,
the same (for useful pedagogical examples of this
statement see, eg., Refs.~\onlinecite{newref1,newref2}).
Thus, the off-shell dependence of the $\rho^0$-$\omega^0$
element of the vector meson propagator matrix,
\begin{eqnarray}
\Delta^{\rho\omega}_{\mu\nu}(q^2)&\equiv&i\int\, d^4q \exp (iq.x)
\, <0\vert T(\rho^0_\mu (x)\omega^0_\nu (0))\vert 0> \nonumber \\
&&\equiv \left( g_{\mu\nu}-{q_\mu q_\nu\over q^2}\right)
\Delta^{\rho\omega}(q^2) \nonumber \\
&&\equiv \left( g_{\mu\nu}-{q_\mu q_\nu\over q^2}\right)
{\theta (q^2)\over (q^2-m^2_\rho )(q^2-m^2_\omega )}\eqnum{1}
\label{one}
\end{eqnarray}
will, in general, be changed when one makes a new choice
of $\rho^0$, $\omega^0$ interpolating fields.
One may readily display interpolating field
choices for which $\theta (q^2)$ necessarily vanishes at
$q^2=0$\onlinecite{ref18} (and hence is obviously $q^2$-dependent),
eg.,
\begin{eqnarray}
&&\rho^0_\mu ={g_\rho\over\hat{m}^2_\rho}V^\rho_\mu\nonumber \\
&&\omega^0_\mu ={g_\omega\over\hat{m}^2_\omega}V^\omega_\mu
\eqnum{2}\label{two}\end{eqnarray}
where $\hat{m}_{\rho ,\omega}$ are the $\rho$, $\omega$ masses,
$V^\rho_\mu =(\bar{u}\gamma_\mu u-\bar{d}\gamma_\mu d)/2$,
$V^\omega_\mu =(\bar{u}\gamma_\mu u+\bar{d}\gamma_\mu d)/6$, and
$g_\rho$, $g_\omega$ are the usual vector meson decay constants,
defined by $<0\vert V^{\rho ,\omega}_\mu \vert \rho ,\omega (q,
\epsilon^\lambda )>\equiv \hat{m}^2_{\rho ,\omega}\epsilon^\lambda_\mu /
g_{\rho ,\omega}$, with $\epsilon^\lambda$ the polarization vector.
However, given the freedom of field redefinition,
this is not enough to exclude the
possibility that, for some other field choice, $\theta (q^2)$
might turn out to be $q^2$-independent.  The field redefinitions
necessary to produce this effect
would then shift the $q^2$-dependence from the propagator
into the
vertices in the new effective Lagrangian.
As pointed out by Cohen and Miller\onlinecite{ref26}, this raises
the possibility that the standard treatment described above might
simply correspond to a different interpolating field choice
than those of the other treatments, one for which $\theta (q^2)$
is $q^2$-independent and the CSV vertices, simultaneously, happen
to approximately vanish.  There is, at present, nothing to rule
out this scenario.
Given the wide range of field choices made possible by the freedom
of field redefinition, it would seem unlikely that one could
make further progress.  However, we will see below that one can in
fact show that (1) there exist interpolating field choices for the vector
mesons for which the full, single vector meson exchange contribution to
NN CSV vanishes at $q^2=0$ and (2)
certain general constraints,
associated with unitarity and analyticity, which must be satisfied for all
choices of interpolating field,
exclude the possibility of finding interpolating fields for
which $\theta (q^2)$
is constant.

In order to set the context for the first point above, it is
useful to begin with a slightly generalized form of an observation
first made by Cohen and Miller.  This states that there is no
{\it algebraic} distinction between CSV NN interactions (at the
one boson exchange level) associated with CSV-vertex and CSV-propagator
contributions.  The argument required to arrive at this observation
runs as follows.  To first order in
isospin breaking, the CSV contributions to NN scattering associated with
vector meson exchange are of two types:  (1) those
involving one charge symmetry conserving (CSC) and one CSV
$\rho^0 NN$ (or $\omega^0 NN$) vertex, combined with a (CSC)
$\rho^0$ (or $\omega^0$) propagator, and (2) those involving
CSC $\rho^0 NN$ and $\omega^0 NN$ vertices together with the CSV
off-diagonal $\rho^0\omega^0$ element of the vector meson propagator.
If we consider the latter contribution, it is (suppressing the Lorentz
indices, $\gamma$ matrices and nucleon spinors, which are inessential
to the argument), of the form
\begin{equation}
v^{(1)}_\tau (q^2) {\theta (q^2)\over (q^2-m^2_\rho )(q^2-m^2_\omega )}
v^{(2)}(q^2)\eqnum{3}\label{three}\end{equation}
where $v^{(1)}_\tau$ is the CSC (isovector) $\rho^0 NN$ and
$v^{(2)}$ the CSC (isoscalar) $\omega^0 NN$ vertex, and the
$q^2$-dependence of the vertices results from
phenomenological form factors, which are supposed to provide
a representation of higher order effects in the effective
meson-baryon theory.  Now imagine that we write
$\theta (q^2) =c+[\theta (q^2)-c]$ and use the standard
partial fraction decomposition
\begin{equation}
{1\over (q^2-m^2_\rho )(q^2-m^2_\omega )}=
{1\over (m^2_\omega -m^2_\rho )}\left[ {1\over q^2-m^2_\omega }
-{1\over q^2-m^2_\rho }\right]\ .\eqnum{4}\label{four}
\end{equation}
The expression (3) can then be re-written as
\begin{eqnarray}
v^{(1)}_\tau (q^2)&& {c\over (q^2-m^2_\rho )(q^2-m^2_\omega )}v^{(2)} (q^2)
+v^{(1)}_\tau (q^2) {b(q^2)\over (q^2-m^2_\omega )}v^{(2)} (q^2)\nonumber \\
&&\qquad\qquad -v^{(1)}_\tau (q^2) {b(q^2)\over (q^2-m^2_\rho )}v^{(2)} (q^2)
\eqnum{5}\label{five}\end{eqnarray}
where
\begin{equation}
b(q^2)\equiv {\theta (q^2) -c\over m^2_\omega -m^2_\rho }\ .
\eqnum{6}\label{six}\end{equation}
The interesting observation made by Cohen and Miller
is that the expression (5) is of precisely
the same form as would result from a combination of three contributions:
(1) a CSV mixed $\rho^0\omega^0$
exchange having constant $\theta (q^2)=c$ and CSC vertices, (2) a CSC
$\omega^0$ exchange
with CSV vertex $v^{(1)}_\tau (q^2)b(q^2)$ and CSC vertex
$v^{(2)}(q^2)$, and (3) a CSC $\rho^0$ exchange with CSV vertex
$v^{(2)} (q^2)b(q^2)$ and CSC vertex $v^{(1)}_\tau (q^2)$.
So far this is no more than algebraic manipulation.
It can, however, be used to give physical meaning to the standard
treatment of few-body CSV if,
first, noting that $\theta (m_\omega^2)$ is ``measured'' in
$e^+e^-\rightarrow\pi^+\pi^-$, one considers,
in the language of the discussion above, $c=\theta (m_\omega^2)$,
and, second, having made this choice, is able to argue that the function
$v^{(2)}(q^2)b(q^2)$ can
be interpreted as a {\it physical} $\rho^0NN$ CSV vertex
(for this particular
choice of $c$, the second term in Eqn.~(5) becomes non-singular
at $q^2=m_\omega^2$ and hence is of the form of a background
contribution).  Eqn.~(5),
with $c=\theta (m_\omega^2)$, would then correspond to
an effective theory with constant $\theta (q^2)$, in which
the additional ``vertex-like'' terms in (5) are viewed as being only a
part of the full contributions associated with the CSV
$\rho^0 NN$, $\omega^0 NN$ vertices of the theory.  If, combined
with the other CSV vertex terms, the net CSV vertex contributions
were to be small, one would then have recovered the standard
treatment.  Given the phenomenological successes of this
approach, it is then suggested that alternate approaches
which find large $q^2$-dependence of $\theta (q^2)$ might, if
they evaluated the CSV vertices using the same choice of
fields, find that the vertex contributions essentially
cancelled the effects of the $q^2$-dependence of $\theta (q^2)$,
restoring the standard treatment.  We now show, however, that
this cannot, in general, be true by giving an explicit choice
of $\rho^0$, $\omega^0$ interpolating fields for which
both $\theta (q^2)$ and the CSV vertices
vanish at $q^2=0$.

Let us consider the simplest (and most natural) field choices for the
$\rho^0$ and $\omega^0$, given by Eqns.~(2).  The off-diagonal
element of the vector meson propagator matrix then becomes, up
to a constant, just the current correlator $<0\vert T(V^\rho_\mu V^\omega_\nu
)\vert 0>$, for which $\theta (q^2)=0$\onlinecite{ref18}.
$V^\rho_\mu$, moreover, is just the third component of the isospin current,
and $V^\omega_\nu$ a linear combination of the hypercharge and baryon
number currents, and all three of these currents remain exactly
conserved, even in the presence of isospin breaking.  As such,
the nucleon matrix elements of these currents are uniquely determined
at $q^2=0$ by the $I_3$, $Y$ and $B$ values of the nucleon.  In
particular, there are no contributions to the $q^2=0$ values of
$<N^\prime \vert V^\rho_\mu\vert N>$ and
$<N^\prime \vert V^\omega_\nu\vert N>$
at any order in $(m_d-m_u)$ or $\alpha_{\rm EM}$, and
hence no CSV vertex contributions at $q^2=0$.  The full NN CSV
contribution due to single $\rho^0$, $\omega^0$ vector meson exchange
graphs of the theory having these interpolating fields thus vanishes
at $q^2=0$, in contrast to the non-zero contribution obtained in the standard
treatment.
All CSV at $q^2=0$ in this case would then have to be associated with multiple
meson and/or heavier meson exchanges
and the full single-vector-meson-exchange CSV contribution
would actually change sign in going from the timelike
to the spacelike region.

It should be stressed that the argument above does not imply that
all NN CSV vanishes at $q^2=0$, only that associated with
single $\rho^0$, $\omega^0$ exchange.
One must, moreover, bear in mind that
it is only the full S-matrix, and not the one-boson-exchange contribution
thereto,
which is independent of interpolating field
choice.  As such, one might still entertain the weaker hypothesis that,
in spite of the above behavior for the interpolating field choice
of Eqns.~(2), there exists some other field choice for which
the standard scenario is realized.
This brings us to our second point, which is to show that even this weaker
hypothesis is untenable.

To begin, let us be more specific
about what would be required to give the algebraic manipulation
above a physical, as opposed to simply
a phenomenological, meaning, namely that
there should be {\it some} choice of $\rho^0$, $\omega^0$
interpolating fields for which the contributions having the
algebraic form of either the CSV propagator mixing or CSV vertex terms
in (5) are actually generated by CSV in the vector meson
propagator matrix or vector meson-nucleon
vertices of the corresponding effective theory
$L_{\rm eff}[\rho^0,\omega^0,\cdots ]$.  Although, as first noted by
Cohen and Miller, the
manipulation above demonstrates that there is no {\it algebraic} distinction
between such sources of CSV, the distinction becomes crucial
once one wishes to use information
extracted from $e^+e^-\rightarrow\pi^+\pi^-$
in the few-body context, since only the CSV-propagator,
and not the CSV-vertex,
contributions are present, and hence (potentially) extractable from the
$e^+e^-\rightarrow\pi^+\pi^-$ experiment.
Although, given the freedom of field
redefinition, a huge class of possible $L_{\rm eff}$'s exists,
and one might despair of obtaining any general information,
valid for all of them, it is presumably the case that,
regardless of interpolating field choice, certain general
properties, such as unitarity, analyticity and the existence of
a spectral representation of the propagator, must be
common to all of them.  As we will now see, this constraint is
enough to rule out the possibility that there exists any
choice of interpolating fields for which $\theta (q^2)$ is
constant.  This is our second result.

Let us consider the off-diagonal element of the
scalar propagator function, $\Delta^{\rho\omega}(q^2)$, defined in Eqn.~(1).
The general form of
the corresponding spectral function is known:  it has poles at
$q^2=m_\rho^2,\, m_\omega^2$, where
$m_\rho^2=\hat{m}_\rho^2-i\Gamma_\rho\hat{m}_\rho$
and $m_\omega^2=\hat{m}_\omega^2-i\Gamma_\omega\hat{m}_\omega$,
and cuts along the positive, real $q^2$ axis, the first of
these beginning at $q^2=4m_\pi^2$.  $\Delta^{\rho\omega}(q^2)$
is then real for $q^2$ real and $<4m_\pi^2$.
In what follows we will ignore the width of
the $\omega$.

Recall that, from the definition of $\theta (q^2)$ in Eqn.~(1),
one has
\begin{equation}
\theta (q^2)=(q^2-m_\rho^2)(q^2-m_\omega^2)
\Delta^{\rho\omega}(q^2)\eqnum{9}\label{nine}
\end{equation}
where the vector meson squared-masses are the {\it complex} pole
positions.  It is crucial to use this form, rather than that
in which the complex pole positions have been replaced by their real
parts, in order to make contact with
the data from $e^+e^-\rightarrow\pi^+\pi^-$~, since
the extracted value of $\theta$
is obtained using a fitting form in which the complex pole
locations are explicitly present in the resonant
denominators (see Ref.~\onlinecite{ref29}).
One can now easily see that $\theta (q^2)$ cannot possibly
be constant, at least below $q^2=4m_\pi^2$.  Indeed, taking
the ratio of Eqn. (7) at two different below-threshold values,
one has
\begin{equation}
{\theta (q_1^2)\over \theta (q_2^2)}=
\left[ {q_1^2-m_\rho^2\over q_2^2-m_\rho^2}\right]
\left[ {(q_1^2-m_\omega^2)\Delta^{\rho\omega}(q_1^2)\over
(q_2^2-m_\omega^2)\Delta^{\rho\omega}(q_2^2)}\right]\ .
\eqnum{8}\label{eleven}\end{equation}
The second factor on the RHS of Eqn.~(8) is (neglecting the
$\omega$ width) real, the first term complex when
$q_1^2\not= q_2^2$.
Thus $\theta (q_1^2)\not= \theta (q_2^2)$ for $q_1^2\not= q_2^2$ and both
below threshold:  constancy of $\theta (q^2)$ below threshold
is incompatible with the required spectral behavior of the
propagator.

One can, in fact, use Eqn.~(8), together with the reality of
$\Delta^{\rho\omega}(q^2)$ below threshold,
to put an ``analyticity constraint''
on the $q^2$-variation of $\theta (q^2)$, i.e. to give a lower
bound on the $q^2$-variation of $\theta (q^2)$ below threshold
compatible with the constraints of
unitarity and analyticity on the spectral function.  Let us re-write
Eqn.~(8) as
\begin{equation}
{\theta (q_1^2)\over \theta (q_2^2)}-1=r(q_1^2,q_2^2)\left[
q_1^2-m_\rho^2\over q_2^2-m_\rho^2\right] -1\eqnum{9}\label{thirteen}
\end{equation}
where, as we have seen,
$r(q_1^2,q_2^2)$ is real for $q_1^2$,$q_2^2$$<4m_\pi^2$,
if we ignore the $\omega$ width.  One may then ask, if one is
allowed to adjust the form of $\Delta^{\rho\omega}(q^2)$
(and hence $r(q_1^2,q_2^2)$) so as to minimize the magnitude
of the RHS of Eqn.~(9), how small can the $q^2$-dependence
be made, subject only to the constraint that $r(q_1^2,q_2^2)$
remain real?
Note that there is no guarantee that one could actually succeed
in finding interpolating fields which realize this lower bound,
given that the requirement corresponds to a highly restrictive
statement about the form and magnitude of the off-diagonal
element of the propagator, and as a result, the actual
$q^2$-variations will, in general, be larger (probably much larger)
than those obtained following the procedure just described.
What one, however, is guaranteed, is that the actual variation
with $q^2$ for {\it any} choice of interpolating fields must be
greater than that specified by the bound so obtained.  Taking the
magnitudes of both sides of Eqn.~(9), and determining the (real)
value of $r(q_1^2,q_2^2)$ which minimizes this magnitude,
one finds,
for example, a minimum variation in the magnitude of
$\theta (q^2)$ of $15\%$ between
$q^2=-1\, $GeV and $q^2=0$.  We stress again that actually reducing
the variation to this level is not necessarily possible, and
that typical variations (as in the case of the vector current
interpolating field choice) will be much greater.  Moreover, although
the argument above can no longer be implemented for $q^2>4m_\pi^2$,
the fact that $q^2$-variation is unavoidable below $q^2=4m_\pi^2$
clearly argues for the likelihood of its presence above this point.

The validity of the argument above, of course, rests on the
fact that the $\rho$ is not a narrow resonance.  If, instead,
both the $\rho$ and $\omega$ had essentially zero width, then
the conclusion could be completely evaded, as is evident
from Eqn.~(8).  The significant effect
of including the width of the $\rho$, has also been stressed recently
in Ref.\onlinecite{ref22}, where it is demonstrated that
taking the spectral function of $\Delta^{\rho\omega}$ to
consist of a sum of constant multiples of the $\rho$ and $\omega$
Breit-Wigner resonance forms leads to significant $q^2$-dependence
of $\Delta^{\rho\omega}$.
While such a form for the spectral function is not the most general
that would be produced if one considered all possible field redefinitions,
(only S-matrix properties, like the pole positions, are
independent of the interpolating field choice),
it should be noted that the effect on the $q^2$-dependence
of including the $\rho$ width is numerically very large.

Another alternative for evading the
argument would be to write, instead of (7),
\begin{equation}
\theta^\prime (q^2)\equiv (q^2-\hat{m}_\rho^2)(q^2-\hat{m}_\omega^2)
\Delta^{\rho\omega}(q^2)\eqnum{10}\label{twelve}\end{equation}
and assume $\theta^\prime (q^2)$ was constant,  which
is then consistent with $\Delta^{\rho\omega}(q^2)$ being purely
real below threshold.  Now, however, one can no longer obtain $\theta^\prime$
from $e^+e^-\rightarrow\pi^+\pi^-$
(which is analyzed using the alternate form having the
correct pole locations).  $\theta^\prime (q^2)$, moreover,
has a pole at $q^2=m_\rho^2$ and a zero at $q^2=\hat{m}_\rho^2$,
and hence is certainly not constant near $q^2=\hat{m}_\rho^2$.  The
resulting rapid variation in the vicinity of $q^2=\hat{m}_\rho^2$
would also make the connection of the value of $\theta$ measured
experimentally (even assuming direct $\omega\rightarrow\pi\pi$ contributions
{\it can} be neglected) to the values of $\theta^\prime (q^2)$ for
$q^2$ on the real axis below the position of the zero, rather
problematic. Finally, one easily sees that, even ignoring the
existence of a nearby pole and zero,
$\theta^\prime$ cannot be constant over the
whole of the range required to save the ``standard'' treatment,
since $\Delta^{\rho\omega}$
has a non-zero imaginary part above $q^2=4m_\pi^2$,
presumably significantly so in the region of the $\rho$ peak,
and this means that,
if $\theta^\prime$ were constant with $q^2$, $\Delta^{\rho\omega}$
would then also have a non-zero imaginary part, eg. at $q^2=0$,
incompatible with the requirement that it be real below
threshold.  This argument can
be evaded, again, only when both resonances have zero width (in
which case the $\rho$ pole moves up to the real axis,
cancelling the factor of $(q^2-\hat{m}_\rho^2)$,
producing a non-zero real part of $\theta^\prime$ and
an imaginary part which vanishes for all
$q^2$, assuming all spectral strength to be isolated at the $\rho$
and $\omega$ poles).

In light of the argument above, it appears that the standard
approach to few-body CSV cannot be physically justified.
While one might worry about the fate of the phenomenological
successes associated with the standard treatment, it is worth
noting that, first, this success is based on the potentially
dangerous assumption of the neglect of
direct $\omega^0\rightarrow \pi\pi$
contributions to $e^+e^-\rightarrow\pi^+\pi^-$ and, second, a
$\rho^0\omega^0$ mixing contribution that is, for example,
half the size of that usually employed would, in fact,
create no phenomenological problems, except for a somewhat
smaller than required non-Coulombic contribution to the $A=3$
binding energy difference.

Concerning the first point, it
should be stressed that, although direct $\omega^0\rightarrow\pi\pi$
contributions to $e^+e^-\rightarrow\pi^+\pi^-$ are usually neglected,
there is actually no good reason to assume that they will
be negligible relative to those associated with $\rho^0\omega^0$
mixing.  Indeed, both are,
in general, non-zero at
${\cal O}(m_d-m_u)$, and should, therefore,
barring other information, be expected to
be comparable in magnitude, as is born out by both a
recent QCD sum rule analysis of the vector current correlator
$<0\vert T(V^\rho_\mu V^\omega_\nu )\vert 0>$\onlinecite{ref21}
and a recent calculation of the direct
$\omega^0\rightarrow\pi\pi$ contribution in a model using confining quark
propagators and Bethe-Salpeter meson-quark vertices
motivated by non-perturbative
Schwinger-Dyson equation studies\onlinecite{ref27}.
Without the neglect of direct $\omega^0\rightarrow\pi\pi$
contributions, however, $e^+e^-\rightarrow\pi^+\pi^-$
cannot be used to give direct information on $\Delta^{\rho\omega}_{\mu\nu}$.
Significant direct $\omega^0\rightarrow\pi\pi$ contributions,
as suggested by the studies mentioned above, would then, of course, call
the phenomenological successes of the standard approach in question.

Concerning the second point (the non-Coulombic contributions to the
$A=3$ binding energy difference), it should be noted that, as pointed
out in Refs.~\onlinecite{ref30,ref31,ref32,ref33},
it is likely that electromagnetic effects associated with
photon-loop, rather than photon-exchange graphs (the former
cannot be disentangled from the strong-interaction, QCD effects
in a model-independent fashion) play a significant role in CSV
in few-body systems, just as, in order to satisfy chiral constraints
on the pion electromagnetic self-energies, they must in the
pseudoscalar spectrum\onlinecite{ref30,ref31,ref32,ref33}.

In summary, it has been shown that
(1) there exist interpolating field choices for the vector mesons
for which the single-vector-meson-exchange contribution to
NN CSV vanishes identically at $q^2=0$ and (2) the assumption
of a momentum-independent
$\theta (q^2)$ for the $\rho^0\omega^0$ propagator is
incompatible with the constraints on the spectral
function associated with analyticity and unitarity.
As such, there can be no choice of $\rho^0$, $\omega^0$
interpolating fields for which the standard approach to
few-body CSV is physically realizable.  Since
the standard approach cannot be interpreted as arising
from any effective meson-baryon Lagrangian it must, in consequence,
be interpreted as being purely phenomenological
in nature.  Given the extremely strong
assumptions required to reduce the $q^2$-variation to even the
level of the analyticity bound discussed above, it
seems clear that, if one wishes to neglect the role of these effects
in few-body systems, it will be necessary to demonstrate explicitly
the existence of interpolating
field choices for which the standard
assumption is an acceptable approximation.

\acknowledgments

The hospitality and stimulating atmosphere of the Workshop on
Non-Perturbative Methods in Quantum Field Theory, at the Australian
National University, are gratefully acknowledged.  The author
also wishes to acknowledge useful conversations with Tony Williams,
Peter Tandy, Tony Thomas and Terry Goldman, to thank
Tom Cohen and Peter Tandy for communicating
the results of Refs.~\onlinecite{ref26} and \onlinecite{ref27}
prior to publication,
and to acknowledge the continuing support of the Natural
Sciences and Engineering Research Council of Canada.

\end{document}